\documentclass{elsart}
\usepackage{natbib, graphicx, amsmath,graphics}

\newcommand{\ie}{{\em i.e.}}
\newcommand{\eg}{{\em e.g.}}

\newcommand{\code}[1]{{\ttfamily #1}}
\newcommand{\MM}[1]{{\boldsymbol #1}}

\DeclareMathOperator{\diff}{d}

\begin{document}
  \begin{frontmatter}
    \centering \title{Diagnostic tools for 3D unstructured oceanographic data}
    
    \vspace{-1.0cm}
    
    \author[1]{C.J.\ Cotter \corauthref{cor1}},
    \corauth[cor1]{Corresponding author}
    \ead{colin.cotter@imperial.ac.uk}
    \author[2]{G.J.\ Gorman},
    
    \address[1]{Department of Aeronautics,\\
      Imperial College London, SW7 2AZ, UK}
    
    \address[2]{Applied Modelling and Computation Group,\\
      Department of Earth Science and Engineering,
      \\Imperial College London, SW7 2AZ, UK}
    
    \journal{Ocean Modelling}
    
    \begin{abstract}
      Most ocean models in current use are built upon structured
      meshes. It follows that most existing tools for extracting
      diagnostic quantities (volume and surface integrals, for
      example) from ocean model output are constructed using
      techniques and software tools which assume structured
      meshes. The greater complexity inherent in unstructured meshes
      (especially fully unstructured grids which are unstructured in
      the vertical as well as the horizontal direction) has left some
      oceanographers, accustomed to traditional methods, unclear on
      how to calculate diagnostics on these meshes. In this paper we
      show that tools for extracting diagnostic data from the new
      generation of unstructured ocean models can be constructed with
      relative ease using open source software. Higher level languages
      such as Python, in conjunction with packages such as NumPy,
      SciPy, VTK and MayaVi, provide many of the high-level primitives
      needed to perform 3D visualisation and evaluate diagnostic
      quantities, \eg\ density fluxes. We demonstrate this in the
      particular case of calculating flux of vector fields through
      isosurfaces, using flow data obtained from the unstructured mesh
      finite element ocean code ICOM, however this tool can be applied
      to model output from any unstructured grid ocean code.
    \end{abstract} 
  \end{frontmatter}
  
  \section{Introduction}
  The development of an ocean model poses a broad range of challenges
  (in addition to the key challenges posed by validation against real
  world circulation and hydrography). Challenges involving the core of
  a model include the design of discretisation schemes, issues of
  numerical stability, good representation of hydrostatic and
  geostrophic balance, data I/O, time varying boundary forcing and
  relaxation to climatology, scalability for parallel computation
  \emph{etc}. In addition to this, models are generally viewed as
  requiring a pre-processor for model preparation (\eg\ mesh
  generation, setting model parameters and boundary conditions), and a
  post-processor for the analysis of model results. Because of their
  maturity, there has been a level of convergence of technologies and
  standards for structured grid ocean models. For example, and the
  subject of this paper, the ocean modelling community has amassed a
  wealth of methods and tools for the analysis of model results:
  Ncview\footnote{Ncview:
    http://meteora.ucsd.edu/~pierce/ncview\_home\_page.html} provides
  a quick and easy way to browse data conforming to the NetCDF Climate
  and Forecast (CF) Metadata
  Convention\footnote{http://www.cfconventions.org/};
  Ferret\footnote{http://ferret.wrc.noaa.gov/Ferret/} is a powerful
  visualization and analysis environment for large and complex gridded
  data sets, supporting numerous gridded file formats and standards
  including OPeNDAP (Open-source Project for a Network Data Access
  Protocol); MATLAB\footnote{http://www.mathworks.com/} is also widely
  used since it combines easily accessible linear algebra routines
  together with interactive graphical output. However, file formats and
  standards for unstructured grid models are only emerging within the
  oceanographic community. For example, only recently has the
  community outlined what a standard might be and a set of milestones
  for implementation (\cite{Aikman06}). Inevitably, application
  programming interfaces (APIs) for these emerging standards are
  still some way off. This in itself is an important consideration for
  researchers interested in developing software tools for analysing
  the output of unstructured ocean mesh models since there is a risk
  that software developed will rapidly become redundant. In addition,
  the actual data is more complex: for example, interpolating a single
  point within an unstructured data set is much more expensive (and
  complex if performed efficiently) than with a simple gridded data
  set which has in general an implicit spatial-temporal index.

  
  Fortunately, there is a rich selection of open source tools which
  facilitates the design and development of diagnostics for complex
  diagnostics on unstructured grids. Diagnostic tools must be
  effective and relatively cheap to create (read scientist
  sweat-and-tears) as the underlying technology is evolving rapidly
  and tools are likely to have a short shelf life. Here we will
  consider the use of Python\footnote{Python: http://www.python.org/}
  (a portable interrupted language) which has risen to prominence in
  science and engineering in recent years, particularly due to the
  addition of libraries such as NumPy and SciPy
  \citep{Oliphant07}. The Visualization ToolKit
  (VTK\footnote{Visualization ToolKit: http://www.vtk.org}) is another
  open source project which provides a Python API (also provides an
  API for C++, Java and TCL/TK), thus enabling a rich environment for
  creating 3D OpenGL scientific visualisation. This API provides ample
  functionality to perform differentiation, integration and
  interpolation on structured grids and unstructured grids containing
  linear tetrahedra, hexahedra, triangular prisms (wedges), pyramids
  and quadratic tethahedra and hexahedra elements as well as linear
  and quadratic triangles and quadrilaterals. The wedge elements are
  used by many ocean models which are unstructured in the horizontal
  but structured in the vertical: this includes SUNTANS (Stanford)
  which is a finite volume code, and SLIM (Louvain-la-Neuve) which
  uses nonconforming and discontinuous linear elements. FEOM
  (Bremerhaven) uses continuous linear tetrahedral elements (with
  wedge elements under development). ICOM (Imperial College London)
  uses tetrahedral and hexahedral elements. Multilayered 2D shallow-water
  models such as Delfin (Delft) and ADCIRC (Notre Dame) could also
  make use of this framework.

  The reason that we choose the VTK/Mayavi combination is that the
  Python model of development, in which the developer time is
  considered at a premium with optimisation taking place only where it
  is necessary, facilates quick development of new tools in a rapidly
  changing scientific environment. Additionally, these projects have
  open source licenses which facilitates cross-project collaboration
  and comparison between models.

  To illustrate what is involved in creating new and novel diagnostics
  for data on unstructured meshes, we extend the open source
  visualisation package MayaVi\footnote{MayaVi:
    http://mayavi.sourceforge.net} \citep{Ra2001}, which is developed
  using Python and VTK. Importantly, diagnostics constructed in the
  manner can be applied to both structured and unstructured data sets;
  the only effort being to write code to convert from the data format
  to a VTK file (which can be done using the VTK API). This is crucial
  both to the portability of the methods and to comparisons of results
  from different models.

  The motivating example developed here is the case of fluxes through
  isosurfaces since they are among the most complicated diagnostic
  quantities to calculate and visualise. The filter that we discuss
  has been added to the Mayavi source trunk and is freely
  available. It can be applied to output from any of the unstructured
  grid models described above (as well as structured grid models).

  In Section \ref{methodology}, we describe the general methodology of
  using VTK and MayaVi, with reference to the isosurface case as an
  example. In Section \ref{examples}, we show diagnostics calculated
  from ICOM output data: namely, temperature fluxes through vertical
  levels and volume flux through temperature isosurfaces calculated
  from a deep convection test case. Section \ref{summary and outlook}
  provides a summary and outlook.

\section{Methodology}
\label{methodology}

A finite element\footnote{Finite volume may be thought of as a
  specific type of finite element method in this context.}
representation defines field values everywhere in the domain, not just
at the grid points. This means that it is possible to define
isosurfaces and integrals uniquely with respect to that
representation; it is also possible to monitor the values of fields at
any location as the solution evolves in time. If the finite element
representation used is piecewise-linear or piecewise-quadratic then
these calculations can all be performed within the VTK framework.

The VTK API readily facilitates a pipeline programming
paradigm. Operations (such as the contour filter used in the following
section) are performed by creating objects which require a reference
to an input data set and provide a reference to an output data set,
which itself may be passed as input to another operation. This means
that when a property changes further up the pipeline, \eg\ a different
field value is chosen for an isosurface filter, that change is
propagated all the way along the chain. This makes applications
developed in this way very interactive: ideal for scientific analysis
of data.

In MayaVi, operations are divided conceptually into modules and
filters. Modules are used to obtain some mode of graphical
representation of the data \eg\ isosurfaces, flow vectors,
streamlines. Filters are used to manipulate the data in some way, \eg\
to extract the Cartesian components of the velocity field as scalars
which can then be visualised using modules designed for
scalars. MayaVi allows the user to add new filters and modules using
the scripting language Python combined with VTK. These new components
can then be contributed back to the MayaVi project. This illustrates
the collaborative power of open source development.

New filters and modules can be introduced into Mayavi as Python
classes. The location of the files containing these classes may be
added to the Mayavi search path using the Mayavi GUI (for more details
see the Mayavi manual). Each class has an \code{initialize} method
which sets up any GUI objects and calls the function to apply the
filter for the first time. Mayavi uses the Python Tkinter API which 
makes it very easy to attach GUI objects to events which are called
whenever the GUI object is changed \emph{e.g.} reconstructing the
isosurface level each time a slider is moved. Filter classes must have 
methods for setting the input
\begin{verbatim}
def SetInput (self, source):
  ...
\end{verbatim}
and the output
\begin{verbatim}
def GetOutput (self):
  ...
\end{verbatim}
to allow several filters to be applied in a chain.

\section{Examples} 
\label{examples}

In this section we demonstrate the capability of this framework using
the example of flow data from a deep convection experiment. This is a
reproduction of the Jones and Marshall experiment described in
\citep{JoMa1993} produced using the unstructured mesh adaptivity
capabilities of ICOM. The original data structure is an unstructured
tetrahedral grid of dimensions $32km\times32km$ in the horizontal and
$2000m$ in the vertical, with a nodal representation of velocity,
temperature (with the reference value $T_0$ subtracted) and pressure
(with hydrostatic and geostrophic components removed) and with linear
interpolation within each tetrahedral element. We chose a snapshot
taken at time $t=48$ hours which exhibits descending plumes and strong
nonhydrostatic dynamics. 

The equations of motion used in the experiment are the nonhydrostatic 
Boussinesq equations with linear equation of state
\[
b = \gamma g(T-T_0)
\]
where $b$ is the buoyancy, $g$ is the gravitational acceleration, $T$
is the temperature and $T_0$ is a reference background
temperature. Thus, after appropriate scaling, the temperature fluxes
can also be interpreted as buoyancy fluxes or heat fluxes. These
fluxes are important diagnostic quantities for this problem because
they describe the advected transport of buoyancy by convection.

\begin{figure}[t]
  \begin{center}
    \includegraphics[height=0.6\textheight]{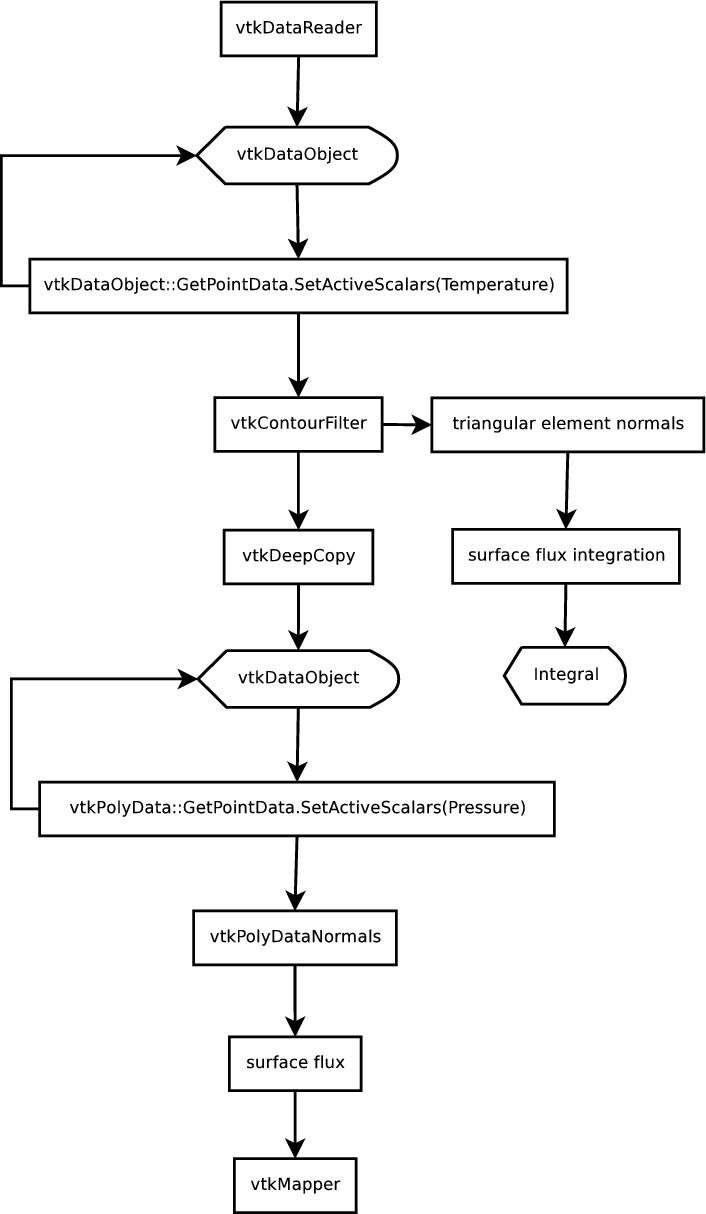}
  \end{center}
  \caption{Data flow for flux visualisation and integration.}
\label{fig:isosurfaceprobe}
\end{figure}

\subsection{Isosurface probe}
To compute these examples we created a new MayaVi filter declared as a class:
\begin{verbatim}
class IsoSurfaceProbe (Base.Objects.Filter):
  ...
\end{verbatim}
 The data flow of this filter is illustrated in Figure
\ref{fig:isosurfaceprobe}. The filter first calculates an isosurface
(using the VTK class \code{vtkContourFilter}) for the active scalar
field (which can be selected using the MayaVi graphical user interface
(GUI)). Regardless of the element type in the input data (meaning that
this filter can be applied to model output from any of the models
described in the introduction), the isosurface itself is comprised of
piecewise-linear triangular elements; any non-triangular polygons are
triangulated (and quadratic triangular elements are split into four
linear triangle elements). The normals obtained are self-consistently
oriented on each connected surface.  Scalars and vectors from the
input data are automatically interpolated using the finite element
basis functions onto the vertices of the isosurface. The contour
filter is set up simply by calling the constructor
\begin{verbatim}
self.cont_fil = vtk.vtkContourFilter ()
\end{verbatim}
The input to the contour filter object is set in the \code{SetInput} method:
\begin{verbatim}
self.cont_fil.SetInput (source.GetOutput ())
\end{verbatim}
For reasons that will become clear in the next section,
we make a deep copy of the output to \code{self.iso}
\begin{verbatim}
self.iso = vtk.vtkPolyData ()
...
self.iso.DeepCopy (self.cont_fil.GetOutput ())
\end{verbatim}
and set the copy as the output to the filter
\begin{verbatim}
def GetOutput (self):
  return self.iso
\end{verbatim} 
so that the data can be visualised.

\begin{figure}
\begin{center}
\includegraphics[height=0.4\textheight]{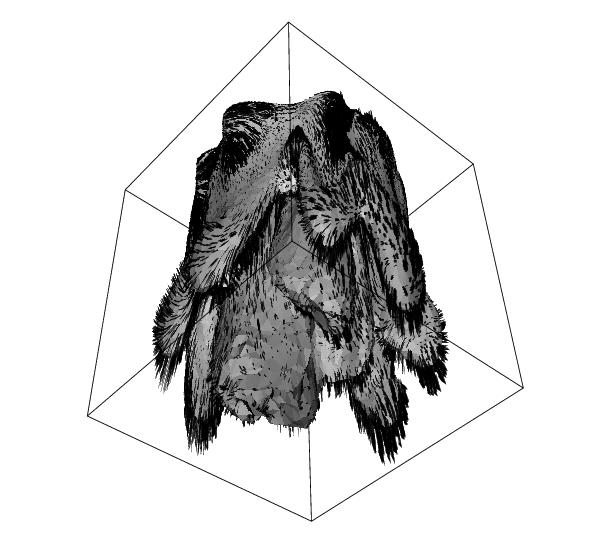} \\
\includegraphics[height=0.4\textheight]{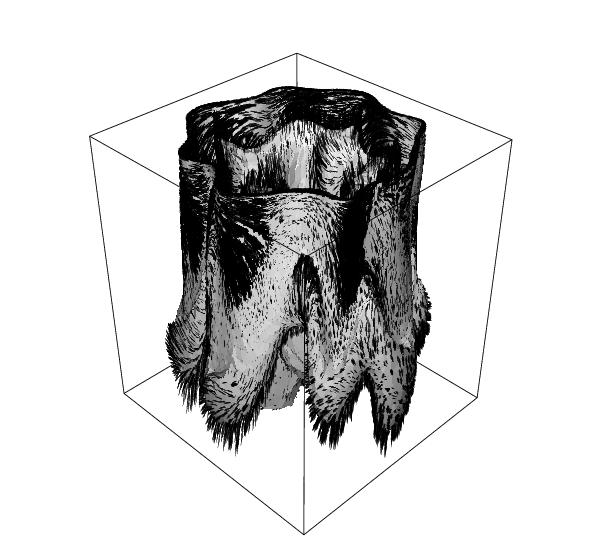}
\end{center}
\caption{Plots showing the temperature $T-T_0=0.02869K$ isosurface,
  together with superimposed velocity vectors projected onto the
  surface, viewed from above and below. The geostrophic rim currents
  are most clearly visible in the view from above, whereas the view
  from below shows the velocity of the rapidly descending plumes.}
\label{isosurface + vecs}
\end{figure}

The output from the filter can be displayed using the standard
MayaVi modules without any further coding. In Figure \ref{isosurface +
  vecs} the surface map module is used to display an isopycnal
surface, with velocity vectors superimposed. This allows the dynamics
of this particular isopycnal surface to be studied. One can see that
the velocity vectors are pointing around the rim of the water column
at the ocean surface in accordance with geostrophic balance, but also
strong downward flow at the head of the descending plumes. See
\citep{JoMa1993} for a description of the dynamical processes.

\subsection{Additional fields}
In many cases it is desirable to add additional diagnostic fields to
the output of the contour filter. However, if the active scalar of the
output from the contour filter is modified, this will actually change
the active scalar of the original data object passed into the contour
filter. For this reason it is necessary to make a copy of the result
of the contour filter (See Figure \ref{fig:isosurfaceprobe}).  With
this copy, new data fields can be added and the active scalar (or
vectors) can be changed. This allows one to visualise arbitrary field
data on the isosurface. 

In our filter, the volume flux $\MM{u}\cdot\MM{n}$ is calculated,
where $\MM{u}$ is the velocity calculated at the isosurface vertices
and $\MM{n}$ is the normal vertices calculated using the VTK class
\code{vtkPolyDataNormals} class. The normals are calculated as points data
on the triangle vertices,
\begin{verbatim}
normal_calculator = vtk.vtkPolyDataNormals()
normal_calculator.SetInput(self.GetOutput())
normal_calculator.Update()
normals = normal_calculator.GetOutput().GetPointData().GetNormals()
\end{verbatim}
and then the volume flux is computed and added to the data set.
\begin{verbatim}
volume_flux_array = vtk.vtkDoubleArray()
volume_flux_array.SetNumberOfTuples(n_points)
volume_flux_array.SetName('VolumeFlux')
vecs = self.GetOutput().GetPointData().GetVectors()
for i in range(n_points):
  (u,v,w) = vecs.GetTuple3(i)
  (n1,n2,n3) = normals.GetTuple3(i)
  volume_flux = u*n1 + v*n2 + w*n3
  volume_flux_array.SetTuple1(i, volume_flux)
self.iso.GetPointData().AddArray(volume_flux_array)
self.iso.GetPointData().SetActiveAttribute('VolumeFlux',0)
\end{verbatim}

\begin{figure}
\begin{center}
\includegraphics[height=0.4\textheight]{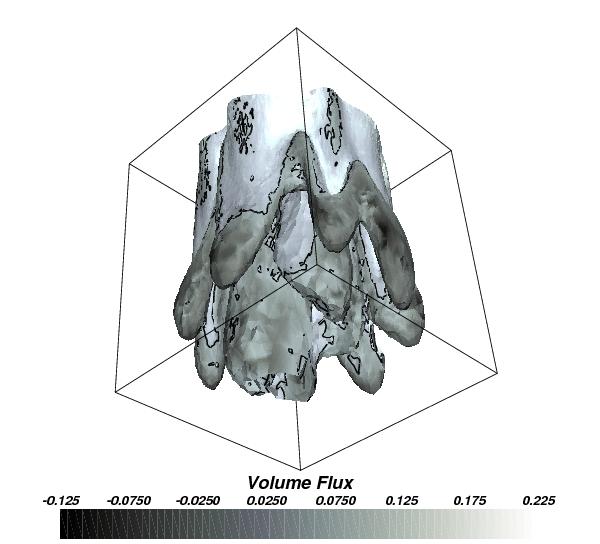} \\
\includegraphics[height=0.4\textheight]{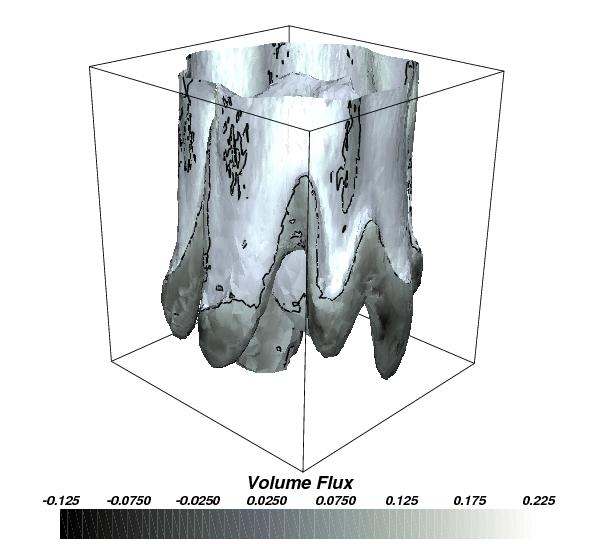}
\end{center}
\caption{Plots showing the temperature $T-T_0=0.02869K$ isosurface
  with a gradient map representing volume flux, viewed from above and
  below. Black lines are used to mark the volume flux zero contour on
  the surface. Negative volume flux indicates the isosurface is
  locally expanding and positive volume flux indicates the isosurface
  is locally contracting (the overall sign depends on how VTK orients
  the surface; this is done self-consistently on each connected
  surface). The plots show that the isosurface is expanding at the
  bottom and shrinking at the top \emph{i.e.} it is being stretched
  out by the flow. This stretching occurs to satisfy incompressibility
  of the flow.}
\label{isosurface + vol flux}
\end{figure}

\begin{figure}
\begin{center}
\includegraphics[height=0.4\textheight]{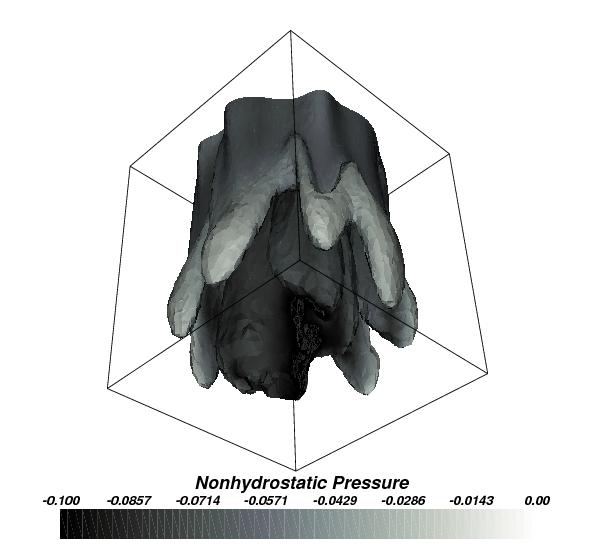} \\
\includegraphics[height=0.4\textheight]{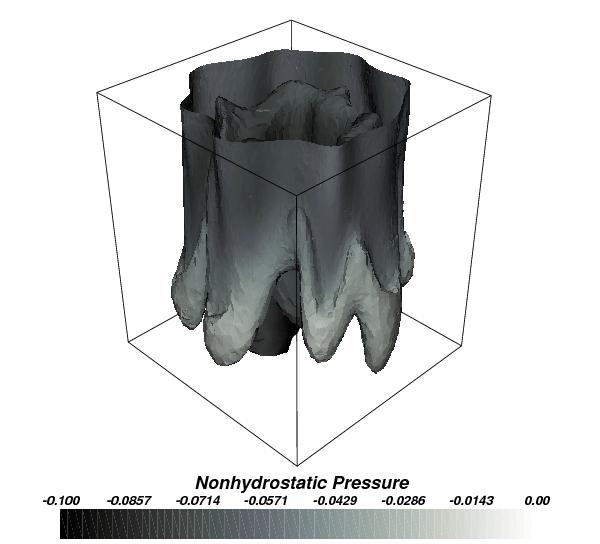}
\end{center}
\caption{Plots showing the temperature $T-T_0=0.02869K$ isosurface with a
  gradient map representing nonhydrostatic pressure, viewed from above
  and below. The nonhydrostatic pressure has the largest magnitude
  around the descending plumes inside the convection cell.}
\label{isosurface + nonhydro}
\end{figure}

The volume flux is shown in Figure \ref{isosurface + vol flux},
visualised using the Mayavi SurfaceMap module together with a zero
contour produced by the IsoSurface module. This field shows where the
isopycnal surface is expanding and where it is contracting, \ie\ it
illustrates how the isopycnal surface is being advected by the
flow. This can be used for diagnosing mixing in the flow as it shows
how small scales are formed in the temperature field. In Figure
\ref{isosurface + nonhydro} a gradient map for nonhydrostatic pressure
is projected onto the same isopycnal to illustrate this. This plot
allows one to study the relative size of nonhydrostatic pressure in
different parts of the convective structure. The plot shows that
nonhydrostatic pressure is greatest near to the plumes at the centre
of the convection cell. Pressure is a global quantity arising from the
pressure Poisson equation so this should not form a complete guide to
nonhydrostatic effects, however it does reveal that there is
significant nonhydrostatic dynamics in the flow.

\begin{figure}
  \centering
  \begin{tabular}{cc}
    \begin{minipage}{0.48\textwidth}
      \includegraphics[width=\textwidth]{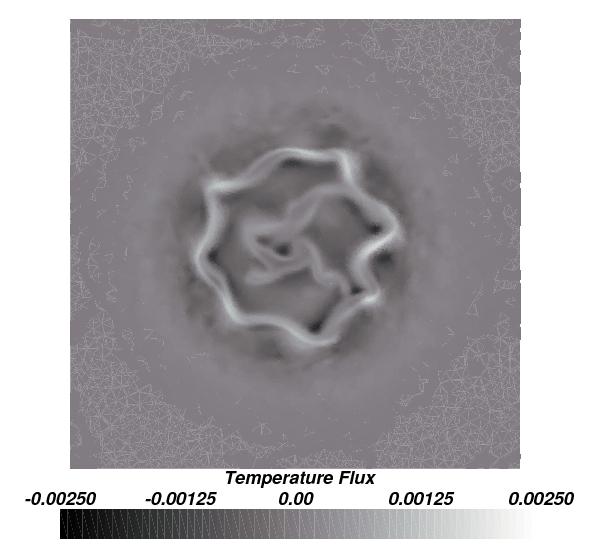}      
    \end{minipage}
    &
    \begin{minipage}{0.48\textwidth}
      \includegraphics[width=\textwidth]{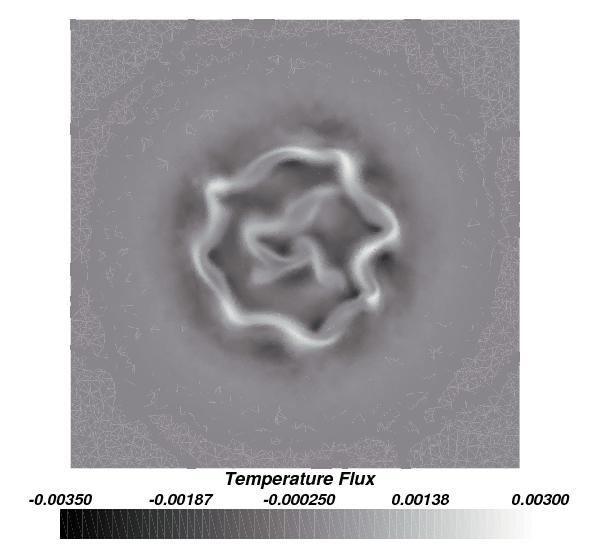}      
    \end{minipage}
    \\
    \begin{minipage}{0.48\textwidth}
      \includegraphics[width=\textwidth]{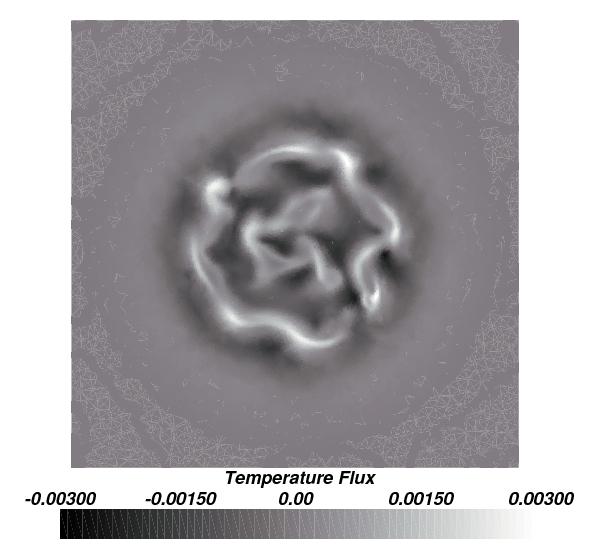}
    \end{minipage}
    &
    \begin{minipage}{0.48\textwidth}
      \includegraphics[width=\textwidth]{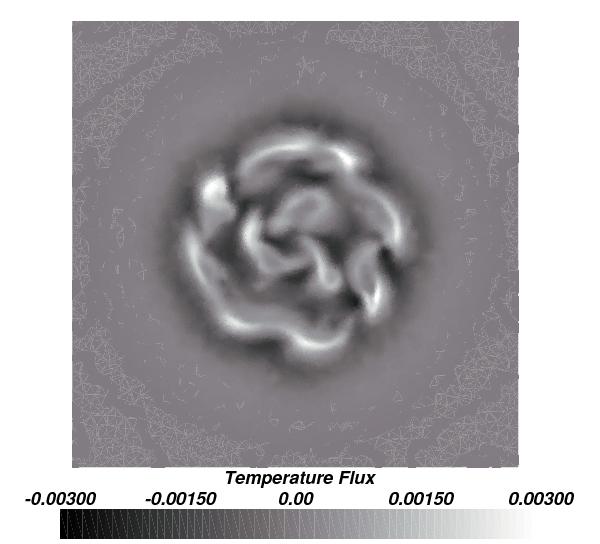}
    \end{minipage}
  \end{tabular}
  \caption{Plots showing temperature flux through horizontal surfaces of
    different levels, defined as isosurfaces of the initial condition
    for temperature. The levels are: $T-T_0 = 0.07$ (top-left),
    $T-T_0=0.065$ (top-right), $T-T_0=0.06$ (bottom-left) and $T-T_0=0.055$
    (bottom-right). These plots illustrate how temperature is being
    transported by the convection cell.}
  \label{temp flux}
\end{figure}

From this point many other diagnostic quantities can readily be
calculated. For example, the temperature advective flux,
$T\MM{u}\cdot\MM{n}$ where $T$ is temperature, through a number of
horizontal slices is shown in Figure \ref{temp flux}. It shows
downward transport of temperature inside the plume structures and a
weak upwelling in-between.

\subsection{Flux calculation}
Mayavi filters and modules can also be used to compute integrals over
the surface. This is done by looping over the triangular faces in the
surface and calculating the contribution from each face using the
piecewise-linear representation. 

For a vector field $\MM{F}$, the flux through an
isosurface $S$ is defined as
\begin{equation*}
  \int_{S} \MM{F} \cdot \MM{n} d S,
\end{equation*}
where $S$ is the isosurface, and $\MM{n}$ is the normal to that
surface. On a piecewise-linear triangular mesh, this is expressed
discretely as:
\begin{equation}\label{eqn:volume_flux}
  \sum_{e=1}^{E} A_e\MM{F}_e \cdot \MM{n}_e,
\end{equation}
where $E$ is the total number of triangular elements defining the
surface, $\MM{F}_e$ is the mean value of $\MM{F}$ on triangle $e$,
$A_e$ is the area of triangle $e$ and $\MM{n}_e$ is the normal to
triangle $e$. 

The flux calculation function in the class makes repeated use of VTK
data retrieval routines.  First it gets \code{vtkDataArray} variables
which point to the active vectors and scalars.
\begin{verbatim}
def calc_flux (self, event=None):
  ...
  point_to_cell = vtk.vtkPointDataToCellData ()
  point_to_cell.SetInput (self.GetOutput ())
  point_to_cell.Update ()
  vecs = point_to_cell.GetOutput ().GetCellData ().GetVectors ()
  fluxf = self.GetOutput().GetCellData().GetScalars()
\end{verbatim}
Next, it loops over the cells (elements) in the isosurface, computes 
the cell area and normals,
\begin{verbatim}
  for cell_no in range(n_cells):
    Cell = self.GetOutput().GetCell(cell_no)
    ...
    Cell_points = Cell.GetPoints ()
    Area = Cell.TriangleArea(Cell_points.GetPoint(0), \
                             Cell_points.GetPoint(1), \
                             Cell_points.GetPoint(2))
    n = [0.0,0.0,0.0]
    Cell.ComputeNormal(Cell_points.GetPoint(0), \
                       Cell_points.GetPoint(1), \
                       Cell_points.GetPoint(2), n)
\end{verbatim}
then it gets the values of the active vectors and scalars in each cell
and computes the volume and scalar fluxes.
\begin{verbatim}
    (u,v,w) = vecs.GetTuple3(cell_no)
    f = Area*(u*n[0] + v*n[1] + w*n[2])
    Integral_volume_flux = Integral_volume_flux + f
    ...
    s = fluxf.GetTuple1(cell_no)
    scalar_advected_flux = scalar_advected_flux + s*f
\end{verbatim}

As a test example, we integrated the flux of velocity (volume flux)
over the temperature isosurface $T-T_0=0.02171Km^{-2}s^{-1}$,
obtaining an integral $I=-3.60\times10^5s^{-1}$. For a divergence-free
vector field we should obtain zero; this computed value is small
compared to the total area ($3.4\times10^8m^2$) of this surface so the
numerical errors from the fluids code and from the integration are
small. We also computed advective integrated temperature fluxes (flux
of $(T-T_0)\MM{u}$) over the levels displayed in Figure \ref{temp
  flux} which are displayed in the following table:
\begin{center}
\begin{tabular}{|c|c|}
\hline 
\begin{tabular}{c}
Initial value \\
of $T-T_0$ \\
\end{tabular}
& 
\begin{tabular}{c}
Temperature flux \\
through
surface \\
\end{tabular}
\\
\hline
$0.055K$ & $-7.14\times10^4Ks^{-1}$ \\
$0.06K$  & $-7.24\times10^4Ks^{-1}$ \\ 
$0.065K$ & $-6.44\times10^4Ks^{-1}$ \\
$0.07K$  & $-3.62\times10^4Ks^{-1}$ \\
\hline
\end{tabular}
\end{center}

\begin{figure}
\begin{center}
\includegraphics[height=0.4\textheight]{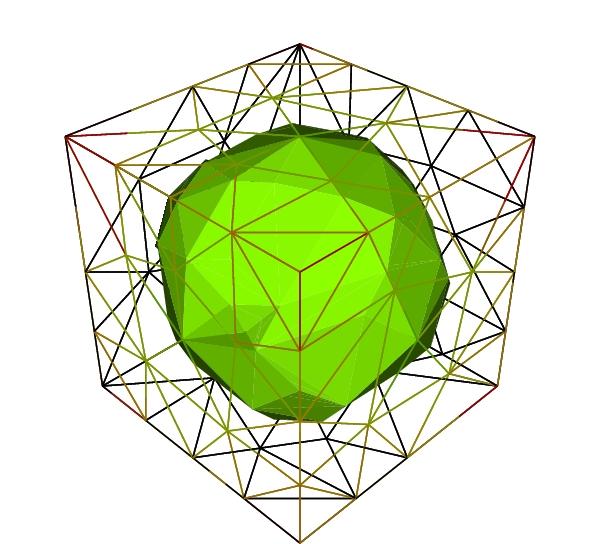}
\includegraphics[height=0.4\textheight]{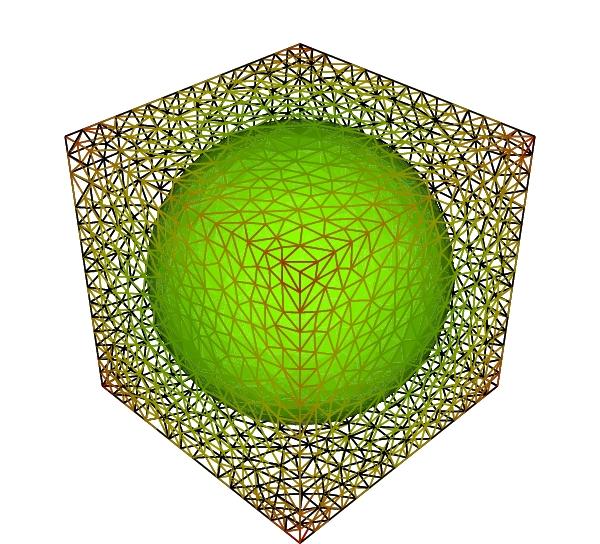}
\end{center}
\caption{\label{cube tests}Plots showing example isosurfaces used to check the
  convergence of flux integrals, with average cell volume 0.01 (top plot)
  and average 0.0001 (top plot).}
\end{figure}

\begin{figure}
\begin{center}
\includegraphics[height=0.4\textheight]{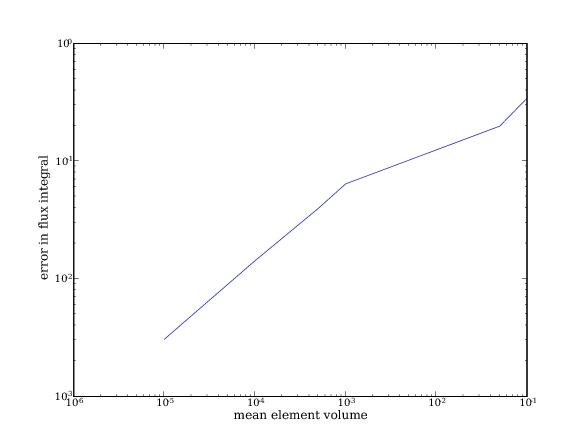}
\end{center}
\caption{\label{cube err}Plot showing error of computed flux against
  average tetrahedral volume for a flux computation on a structured
  mesh.}
\end{figure}

As Mayavi/VTK uses the finite element representation of the solutions
fields (in this paper we restricted ourselves to piecewise-linear
tetrahedral elements), the construction of the isosurface and the
evaluation of flux integrals are exact for the given finite element
representation. This means that the accuracy of the flux integrals are
entirely determined by how well the solution fields are represented on
the mesh. To illustrate this, we took a series of isotropic,
homogeneous, unstructured grids in a cube with dimensions
$1\times1\times1$, and evaluated the field
\[
T = \sqrt{ (X-0.5)^2 + (Y-0.5)^2 + (Z-0.5)^2}
\]
at the grid points. We computed the $T=0.5$ isosurface and calculated
the flux of the vector field 
\[
\MM{F} = (X,Y,Z)
\]
through the isosurface, which has the exact integral 
\[
\int_{T=0.5}\MM{F}\cdot\MM{n}\diff{S}=\pi.
\]
Plots of example isosurfaces are given in Figure \ref{cube tests}. A
plot of the error in the computed flux is given in Figure
\ref{cube err}. We note that whilst these meshes are isotropic and
homogeneous, a more efficient way to ensure that the fields are well-represented on
the mesh is to use anistropic dynamic adaptivity during the calculation
of the flow solution. In this case the accuracy of the flux integrals
(as well as the solution itself) will be determined by the metric used
to construct the mesh.

\section{Summary and Outlook}
\label{summary and outlook}
In this paper we describe a strategy for obtaining diagnostic
information from unstructured adaptive ocean model output by adding
modules and filters to the open source visualisation package MayaVi
using the VTK graphics library. We explained this strategy with the
example of an isosurface probe filter which interpolates flow data
onto an isosurface of a chosen field, and constructs flux quantities
across the isosurface which may then be visualised and integrated. We
illustrated all of this using unstructured flow data from a deep
convection experiment using ICOM.

The strength of the finite element method is that it provides a
representation of the solution fields everywhere in space \emph{via}
the finite element basis functions, \emph{i.e.} not just at the nodal
points where the data is stored. This means that structures such as
isosurfaces are well-defined, as are integrals. Additionally, when the
solution has been evolved using an adaptive unstructured mesh
constructed from an error metric \citep{PaUmOlGo2001} the error in
the integral is bounded by construction.

Python and VTK provides a convenient way to construct new fields and
to calculate integral quantities, and the use of pipelines maximises
the interactivity of any modules and filters added to MayaVi. This
interactivity is an important part of scientific analysis and
exploration of data. Modules and filters are written using the Python
scripting language which means that it is relatively quick to develop
new analysis tools as the need arises without a deep understanding of
the software. The open source environment provides the opportunity for
rapid scientific advance and collaboration.

\section{Acknowledgements}
  This paper began after conversations about diagnostic calculations
  on unstructured grids with Katya Popova. The authors would like to
  acknowledge all the ICOM developers for their collaborative
  support. The deep convection simulation data was provided by Lucy
  Bricheno. This work was partially supported by the NERC RAPID
  Climate Change grant NER/T/S/2002/00459 and the NERC Consortium
  grant NE/C52101X/1. Thanks to Prabhu Ramachandran for developing and
  maintaining the MayaVi project and for responding so helpfully to
  our questions and queries.

  \bibliographystyle{elsart-harv}
  \nocite{PaPiGo2005}
  \bibliography{Fluxes}

\end{document}